\documentclass[prd,twocolumn,nofootinbib,reprint]{revtex4-1}
\usepackage{amsmath,amssymb,amsthm,bm,braket,ascmac,bbm}
\usepackage{graphicx}
\usepackage{color} 
\usepackage{hyperref}

\theoremstyle{definition}

\begin{document}
\title{
Supersymmetric Alignment Models for $(g-2)_{\mu}$
}
\author{Yuichiro Nakai$^1$, 
Matthew Reece$^2$, and Motoo Suzuki$^1$}
\affiliation{\vspace{2mm} \\
$^1$Tsung-Dao Lee Institute and School of Physics and Astronomy,
Shanghai Jiao Tong University, 800 Dongchuan Road, Shanghai, 200240, China\\
 $^2$Department of Physics, Harvard University,
17 Oxford St., Cambridge, MA, 02138, U.S.A
}

\begin{abstract}

Hierarchical masses of quarks and leptons are addressed by imposing horizontal symmetries.
In supersymmetric Standard Models, the same symmetries play a role in suppressing flavor violating processes
induced by supersymmetric particles.
Combining the idea of spontaneous CP violation to control contributions to electric dipole moments (EDMs),
the mass scale of supersymmetric particles can be lowered.
We present supersymmetric models with $U(1)$ horizontal symmetries 
and discuss CP and flavor constraints. 
Models with two $U(1)$ symmetries are found to give a viable solution to the muon $g-2$ anomaly. 
Interestingly, the parameter space to explain the anomaly will be probed by future electron EDM experiments.

 \end{abstract}

\maketitle

%#######################
\section{Introduction}

Precise comparison between theory and experiment 
for the anomalous magnetic moment of the electron led us to the foundation of quantum field theory
to describe the nature of elementary particles. 
In 1947, Kusch and Foley discovered a tiny deviation from $g =2$ in the gyromagnetic ratio of the electron
\cite{PhysRev.72.1256.2},
which was shown to originate from the one-loop effect of QED with the proper treatment of divergences by Schwinger
\cite{PhysRev.73.416}.
Likewise, the reported discrepancy in the anomalous magnetic moment of the muon $(g-2)_\mu$
may lead us to a deeper understanding of nature.
Combining Brookhaven and Fermilab data to measure $(g-2)_\mu$, the discrepancy between theory and experiment is
$\Delta a_\mu^{\rm obs} = a_\mu^{\rm exp} - a_\mu^{\rm theory} = (25.1 \pm 5.9) \times 10^{-10}$
with $a_\mu \equiv (g-2)/2$
\cite{Bennett:2006fi,Keshavarzi:2018mgv,Abi:2021gix}
(for a review of the status of the Standard Model (SM) calculation, see ref.~\cite{Aoyama:2020ynm}\footnote{For an alternative take, see the recent lattice calculation of ref.~\cite{Borsanyi:2020mff}. Intriguingly, if this calculation is correct, the discrepancy with the data-driven calculation of the hadronic vacuum polarization could point to different tensions with the Standard Model~\cite{Lehner:2020crt,Crivellin:2020zul,Keshavarzi:2020bfy,deRafael:2020uif,Malaescu:2020zuc}.}).
Since the discrepancy is of the order of the SM electroweak contribution,
$a_\mu ({\rm EW}) = (15.4 \pm 0.1) \times 10^{-10}$,
it may indicate a new contribution to the muon $g-2$ from physics beyond the SM which lies
at a scale $\lesssim \mathcal{O}(1) \, {\rm TeV}$.

Low-energy supersymmetry (SUSY) has been an attractive candidate of physics beyond the SM to explain the muon $g-2$ anomaly
as well as to address the naturalness problem of the electroweak scale.
Loops of sleptons and electroweakinos can generate a new contribution to the muon $g-2$ with the correct size~\cite{Moroi:1995yh}, an idea that was explored after the first Brookhaven result~\cite{Everett:2001tq,Feng:2001tr,Baltz:2001ts,Chattopadhyay:2001vx,Komine:2001fz} and has continued to draw recent attention (e.g.,~\cite{Iwamoto:2021aaf,Baum:2021qzx,Athron:2021iuf,Baer:2021aax,Ellis:2021zmg}).
However, the required light sleptons and electroweakinos generally lead to dangerously large lepton flavor violation (LFV) in processes such as $\mu \to e \gamma$ and $\mu \to e$ conversion.
To make matters worse, an arbitrary CP violation (CPV) in SUSY breaking parameters
induces a large electric dipole moment (EDM) of the electron,
which is severely constrained by the ACME measurement
\cite{Andreev:2018ayy}
(whose implications for SUSY have been explored in refs.~\cite{Nakai:2016atk,Cesarotti:2018huy}). Even with minimal flavor violation, new physics explaining the muon $g-2$ anomaly with an ${\cal O}(1)$ CPV phase would predict the electron EDM to be five orders of magnitude larger than observations allow. 
Elaborate mechanisms of SUSY breaking such as gauge mediation (see refs.~\cite{Giudice:1998bp,Kitano:2010fa} for reviews)
and gaugino mediation~\cite{Kaplan:1999ac,Chacko:1999mi}
are able to address the issue.

In this paper, we assume the simplest and the most naive version of SUSY breaking without elaborate mechanisms but
introduce $U(1)$ horizontal symmetries,
which can explain the hierarchical masses of quarks and leptons
\cite{Froggatt:1978nt}.
A virtue of such supersymmetric SMs, which we call {\it SUSY alignment models},
is that horizontal symmetries also control the structure of sfermion masses
and suppress flavor violating processes
\cite{Leurer:1992wg,Nir:1993mx,Leurer:1993gy,Ibanez:1994ig}.
Generic CP violation is still dangerous, but
the idea of spontaneous CP violation~\cite{Nir:1996am} makes it possible to suppress contributions to EDMs
and at the same time accommodate the correct Cabibbo-Kobayashi-Maskawa (CKM) phase in the quark sector.
Then, the application of $U(1)$ horizontal symmetries and spontaneous CP violation to the lepton sector
can suppress SUSY contributions to LFV and CPV processes
\cite{Grossman:1995hk,Aloni:2021wzk},
which opens up a possibility to explain the muon $g-2$ anomaly without contradicting CP and flavor constraints.
Our starting point is a model with a single $U(1)$ horizontal symmetry.
We investigate contributions to LFV and CPV processes
and find that the model can achieve supersymmetric particles at around $10 \, \rm TeV$ for a large $\tan \beta$.
Then, we consider models with two $U(1)$ symmetries.
These models can further relax CP and flavor constraints and provide a viable solution to the muon $g-2$ anomaly.
We discuss a relation between the SUSY contribution to the muon $g-2$ and that of the electron EDM
and find that the model parameter space to explain the muon $g-2$ anomaly will be probed by near-future electron EDM experiments.

The rest of the paper is organized as follows. 
In section~\ref{SMflavor}, we review the SM flavor structure which will be explained by $U(1)$ horizontal symmetries.
Section~\ref{Model} presents a SUSY alignment model with a single $U(1)$.
We introduce three different flavon fields which break the $U(1)$ symmetry and CP spontaneously
and discuss LFV and CPV constraints on the model.
In section~\ref{viableModel}, we consider models with two $U(1)$ symmetries
to pursue the possibility of explaining the muon $g-2$ anomaly.
LFV and CPV constraints on these models are investigated in section~\ref{flavorconst}.
Section~\ref{Conclusion} is devoted to conclusions and discussions.
Some details on the presented models and analyses are described in appendices.

%#######################
\section{The SM flavor structure}\label{SMflavor}

The quark mass ratios and mixings in the CKM matrix can be expressed in terms of the Wolfenstein parameterization~\cite{Wolfenstein:1983yz}. The mass ratios are parameterized by powers of $\lambda\sim 0.2$ as
\begin{equation}
\begin{split}
\label{eq:quark_mass}
&m_c/m_t\sim\lambda^3\ , \quad m_u/m_t\sim\lambda^6-\lambda^7,\\[1ex]
&m_b/m_t\sim \lambda^2\ , \quad m_s/m_b\sim \lambda^2\ , \quad m_d/m_b\sim \lambda^4 \ .
\end{split}
\end{equation}
Here, the subscripts of $u,~c,~t,~d,~s,~b$ denote up, charm, top, down, strange and bottom quarks, respectively.
The orders of magnitude of different CKM entries are determined by $\lambda$ as well,
\begin{equation}
\begin{split}
    |V^{\rm CKM}_{12}|\sim \lambda\ , \quad |V^{\rm CKM}_{23}|\sim \lambda^2\ ,
    \quad |V^{\rm CKM}_{13}|\sim \lambda^3\ ,
    \label{ckm}
\end{split}
\end{equation}
where $|V^{\rm CKM}_{ij}|$ is the absolute value of the $(i,j)$ component of the CKM matrix.
The CPV effect in the quark mixings is parameterized by a phase $\delta_{\rm CKM} \simeq 1.2$ in the standard parameterization \cite{Zyla:2020zbs}.

Similarly, the mass ratios and mixing angles of the lepton sector can be expressed in powers of $\lambda$.
The charged lepton mass ratios are written as
\begin{align}
\label{eq:lepton_mass}
    m_\mu/m_\tau\sim \lambda^2\ , \quad m_e/m_\tau\sim \lambda^5\ , \quad m_\tau/m_t\sim \lambda^3\ ,
\end{align}
where subscripts $e,~\mu,~\tau$ denote electron, muon and tau leptons, respectively.
In the present paper, we focus on the normal ordering of neutrino masses with $\nu_3$ ($\nu_1$)
being the heaviest (lightest) neutrino
and for simplicity assume the neutrino mass ratios,
\begin{equation}
\begin{split}
    &m_{\nu_1}/m_{\nu_3}\lesssim \lambda\ , \quad m_{\nu_2}/m_{\nu_3}\sim 1\ .
\label{eq:MNS_matrix}
\end{split}
\end{equation}
Note that a very light or even massless $\nu_1$ is consistent with the neutrino mass-squared difference measurements
and the cosmological constraint on the sum of neutrino masses~\cite{Aghanim:2018eyx}.
The absolute value of the $(i,j)$ component of the Pontecorvo–Maki–Nakagawa–Sakata (PMNS) matrix can be expressed in terms of $\lambda$ as
\begin{equation}
\begin{split}
    &|V^{\rm PMNS}_{12}|\sim \lambda\ , \quad |V^{\rm PMNS}_{13}|\sim \lambda\ , \quad |V^{\rm PMNS}_{23}|\sim 1\ .
    \label{pmns}
\end{split}
\end{equation}
A CPV effect in the lepton mixings parameterized by a phase $\delta_{\rm CP}$
can be measured by the neutrino oscillation,
and $\delta_{\rm CP}\gtrsim \pi$ is favored in the T2K experiment~\cite{Abe:2019vii}. 
The CP-preserving scenario, i.e., $\delta_{\rm CP}=\pi$, is still consistent within the $3\sigma$ confidence level (and the NO$\nu$A experiment has results in mild tension with those of T2K~\cite{Kolupaeva:2020pug}), but future measurements at the Hyper-Kamiokande~\cite{Abe:2018uyc} and DUNE~\cite{Abi:2018dnh} experiments
will provide a more accurate measurement of the lepton sector CPV.

%#######################
\section{The model with a single $U(1)$}\label{Model}

We here assume our supersymmetric SM respects a single $U(1)_H$ horizontal symmetry
under which quark and lepton supermultiplets have nontrivial charges.
Let us first describe a simple example with the $U(1)_H$ symmetry
to motivate a model with three flavon fields.
In the simple model, the fermion mass ratios and mixing matrices are explained by a flavon chiral superfield $S_1$
with charge $-1$ under the $U(1)_H$ symmetry.
The flavon is regarded as a spurious field whose vacuum expectation value (VEV) is given by
$\langle S_1 \rangle = \lambda \Lambda$.
Here, $\Lambda$ denotes some UV mass scale,
and we use $\Lambda=1$ units in the following discussion. 
The $U(1)_H$ charges of quarks and leptons are non-negative integers, and
with appropriate charge assignments, the fermion mass ratios and mixing matrices presented in the previous section are realized.
In particular, the CKM and PMNS mixing matrices are approximately given by 
\begin{align}
\begin{split}
\label{CKMPMNS}
    &|V_{ij}^{\rm CKM}|\sim \lambda^{|H(Q_i)-H(Q_j)|}  \, , \\[1ex]
    &|V_{ij}^{\rm PMNS}|\sim \lambda^{|H(L_i)-H(L_j)|}  \, ,
\end{split}
\end{align}
where $H(X)$ denotes the horizontal $U(1)_H$ charge for a chiral superfield $X$,
and $Q_i$ and $L_i$ $(i=1,2,3)$ are the left-handed doublet quarks and leptons of the $i$-th generation, respectively.
The charges $H(Q_1), H(Q_2), H(Q_3)$ are chosen to realize the CKM entries \eqref{ckm},
and this charge assignment also determines the scaling of the soft scalar mass-squared matrix in terms of $\lambda$,
\begin{align}
   \left. M^2_{\widetilde Q} \right. \sim\widetilde m_q^2
\left(
\begin{array}{ccc}
1 & \lambda &  \lambda^3 \\
 \lambda & 1 & \lambda^2\\
 \lambda^3 & \lambda^2 & 1
\end{array}
\right) ,
\end{align}
where $\widetilde m_q$ denotes a typical mass scale of squarks.
The sizable off-diagonal components in the above matrix lead to dangerously large flavor changing neutral currents (FCNCs).
In particular, 
the $(1,2)$ element is less suppressed compared to the other off-diagonal elements,
and the stringent neutral Kaon mixing constraint requires $\widetilde m_q \gg 10\,{\rm TeV}$.
A similar argument can be applied to the lepton sector.
A reasonable charge assignment for the left-handed doublet leptons to explain the PMNS mixing matrix
leads to sizable off-diagonal elements in the left-handed slepton soft mass-squared matrix,
and then the constraint from the ${\rm Br}(\mu\to e + \gamma)$ measurement requires $\widetilde m_{\ell} \gg 10 \, {\rm TeV}$
for a large $\tan \beta$
where $\widetilde m_\ell$ is a typical mass scale of sleptons.
However, such stringent constraints on the soft mass scales are not a generic consequence of SUSY alignment models.
We will next see that weaker bounds can be obtained by extending the model.

\subsection{Three flavons}

Let us now introduce three flavon fields $S_2, \bar{S}_2, S_3$ with horizontal charges $-2, 2, -3$, respectively.
We suppose these flavons obtain VEVs,
$\langle S_2\rangle\sim \lambda^2$, $\langle \bar S_2\rangle\sim \lambda^2$,~$\langle S_3\rangle\sim \lambda^3$.
As in the case of the simple model above,
appropriate $U(1)_H$ charge assignments will lead to the correct fermion mass ratios and mixing matrices.
However, in the present model, the $(1,2)$ element of $M^2_{\widetilde Q}$ is given by, e.g.,
\begin{align}
    (M^2_{\widetilde Q} )_{12} \sim \langle \bar S_2 \rangle \langle S_3 \rangle
    \widetilde m^2 \sim \lambda^5 \widetilde m^2  ,
\end{align}
which is significantly suppressed compared to that of the simple model.
The same idea is also applied to the lepton sector.

We consider spontaneous CPV to control complex phases in SUSY breaking parameters but still generate (at least) the CKM phase.
A CP invariant superpotential of the flavon fields is given by
\begin{align}  \label{eq:CPVsuper1}
W_S = \xi X(S_2\bar S_2-\lambda^4)+Y(c_1S_3^4+c_2S_3^2 S_2^3+c_3S_2^6) \, ,
\end{align}
where $\xi$ and $c_{1,2,3}$ are real coefficients and 
$X, Y$ are SM gauge singlet chiral superfields but $Y$ has the horizontal charge $+12$. 
The above superpotential respects the $U(1)_R$ symmetry under which $X, Y$ have the charge $+2$ and the three flavons are neutral.
The $F$-term condition for $X$ leads to the VEVs of $\langle S_2\rangle = \langle\bar S_2\rangle = \lambda^2$,
while the condition for $Y$ gives
\begin{align}
c_1 {\langle S_3 \rangle}^4+c_2 {\langle S_3 \rangle}^2 \langle S_2\rangle^3+c_3\langle S_2\rangle^6=0\ ,
\end{align}
which is solved as
\begin{align}
\langle S_3\rangle^2 = \frac{-c_2 \pm
\sqrt{c_2^2 -4c_1 c_3 } }{2c_1} \, \langle S_2\rangle^3\ .
\end{align}
We can see that the VEV $\langle S_3 \rangle$ obtains a complex phase when
\begin{align}
c_2^2 -4c_1 c_3 <0 \ ,
\end{align}
and the CP symmetry is broken spontaneously.

The superpotential for quarks and leptons in the gauge eigenbasis is given by
\begin{eqnarray}
\label{eq:superpotential}
    W_{\rm Yukawa}=&& \!\!\!\!\! Y_{u \,ij} Q_i\bar u_j H_u+Y_{d\,ij} Q_i\bar d_j H_d+Y_{e\,ij} L_i\bar e_j H_d \nonumber \\[1ex]
    &+& Y_{\nu\,ij} \frac{(H_uL_i)(H_u L_j)}{M_N}\ ,
\end{eqnarray}
where $\bar u_i,~\bar d_i,~\bar e_i$ $(i = 1,2,3)$ and $H_{u,d}$ denote the right-handed up-type quarks, the right-handed down-type quarks, the right-handed charged leptons and the two doublet Higgs fields, respectively.
$Y_{u,d,e,\nu}$ are their Yukawa couplings and $M_N$ is some UV mass scale. 
The origin of the Majorana neutrino mass terms in the second line through the seesaw mechanism
\cite{Yanagida:1980xy,Minkowski:1977sc,GellMann:1980vs}
is discussed in Appendix~\ref{app:seesaw}.
To realize the fermion mass ratios and mixing matrices,
we take the following charge assignments for (s)quarks and (s)leptons under the horizontal symmetry, 
\begin{equation}
\begin{split}
\label{eq:single_charge}
&\begin{array}{c}
Q_1 \\
(3)
\end{array}
\begin{array}{c}
Q_2 \\
(2)
\end{array}
\begin{array}{c}
Q_3 \\
(0)
\end{array}
\quad
\begin{array}{c}
\bar{u}_1 \\
(4)
\end{array}
\begin{array}{c}
\bar{u}_2 \\
(1)
\end{array}
\begin{array}{c}
\bar{u}_3 \\
(0)
\end{array}
\quad
\begin{array}{c}
\bar{d}_1 \\
(-7)
\end{array}
\begin{array}{c}
\bar{d}_2 \\
(-4)
\end{array}
\begin{array}{c}
\bar{d}_3 \\
(0)
\end{array} \\[1ex]
&\begin{array}{c}
L_1 \\
(2)
\end{array}
\begin{array}{c}
L_2 \\
(1)
\end{array}
\begin{array}{c}
L_3 \\
(1)
\end{array}
\quad
\begin{array}{c}
\bar{e}_1 \\
(3)
\end{array}
\begin{array}{c}
\bar{e}_2 \\
(-3)
\end{array}
\begin{array}{c}
\bar{e}_3 \\
(-1)
\end{array}
\end{split}
\end{equation} 
We also assume $H(H_u) = H(H_d) = 0$. The above charge assignments work well for a large $\tan\beta\sim 50$
because, e.g., the bottom and tau Yukawa couplings are not suppressed by the horizontal symmetry.

The Yukawa matrices are given by couplings with the three flavons,
and their orders of magnitude can be parameterized by $\lambda$,
\begin{equation}
\begin{split}
\label{eq:q_yukawa_extended}
&Y_u\sim
\left(
\begin{array}{ccc}
\lambda^7 &  \lambda^4 & \lambda^3 \\
\lambda^6  & \lambda^3 & \lambda^2\\
\lambda^4 & \lambda^5 & 1
\end{array}
\right), \qquad Y_d\sim
\left(
\begin{array}{ccc}
\lambda^4 & \lambda^7 & \lambda^3\\
\lambda^{11} &\lambda^2 & \lambda^2\\
\lambda^{13} & \lambda^4 & 1
\end{array}
\right),\\[1ex]
&Y_e\sim
\left(
\begin{array}{ccc}
\lambda^5 & \lambda^7 & \lambda^5\\
\lambda^4 & \lambda^2 & 1\\
\lambda^4 & \lambda^2 & 1
\end{array}
\right), \qquad Y_\nu\sim
\left(
\begin{array}{ccc}
\lambda^4 & \lambda^3 & \lambda^3\\
\lambda^3 & \lambda^2 & \lambda^2 \\
\lambda^3 & \lambda^2 & \lambda^2
\end{array}
\right),
\end{split}
\end{equation}
where an $\mathcal{O}(1)$ coefficient and a possible phase of each matrix element has been omitted for notational simplicity.
The CKM matrix is written in terms of the diagonalization matrices of the left-handed up-type quarks $V_{u}$
and the left-handed down-type quarks $V_{d}$, i.e., $V^{\rm CKM}=V_u^\dagger V_d$.
The up-type quark Yukawa matrix $Y_u$ provides a diagonalization matrix close to the CKM matrix $V_u^\dagger \sim V_{\rm CKM}$,
while $Y_d$ does not provide a sizable mixing between $Q_1$ and $Q_2$ to explain $|V_{12}^{\rm CKM}|\sim \lambda$
because the corresponding off-diagonal entries of $Y_d$ are very small. 
The small $(1,2)$ mixing of $V_d$ makes it possible to suppress SUSY contributions to, e.g.,
the neutral Kaon mixing compared to the simple model.
A similar discussion also applies to the lepton sector,
where the sizable $|V^{\rm PMNS}_{12}|$ is mainly provided via the neutrino Yukawa matrix $Y_\nu$. 
CP phases are introduced through couplings with $S_3$. 
For the CKM phase, for example, the $(1,3)$ entry of $Y_d$ has a phase via the coupling with $S_3$ and the phase can provide $\delta_{\rm CKM}\sim 1$ in a similar manner to that discussed in ref.~\cite{Nir:1996am}.

The structure of the soft scalar mass-squared matrices are constrained as
\begin{equation}
\begin{split}
\label{eq:squark_msq} 
&M^2_{\widetilde Q}\sim\widetilde m_q^2
\left(
\begin{array}{ccc}
1 & \lambda^5 & \lambda^{3} \\
\lambda^5  & 1 & \lambda^2\\
\lambda^3 & \lambda^2 & 1
\end{array}
\right) ,
\quad M^2_{\widetilde {\bar u}}\sim\widetilde m_q^2
\left(
\begin{array}{ccc}
1  & \lambda^3 &  \lambda^4\\
 \lambda^3 & 1 & \lambda^5\\
 \lambda^4 & \lambda^5 & 1
\end{array}
\right),\\[1ex]
&M^2_{\widetilde{\bar d}} \sim \widetilde m_q^2
\left(
\begin{array}{ccc}
1 &  \lambda^3 &\lambda^{7}\\
 \lambda^3 & 1 &\lambda^4\\
\lambda^{7} & \lambda^4 & 1
\end{array}
\right),\\[1ex]
&M^2_{\widetilde L}\sim\widetilde m_{\ell}^2
\left(
\begin{array}{ccc}
1 & \lambda^5 &  \lambda^5 \\
 \lambda^5 & 1 & 1\\
 \lambda^5 & 1 & 1
\end{array}
\right), \quad M^2_{\widetilde{\bar e}}\sim\widetilde m_{\ell}^2
\left(
\begin{array}{ccc}
1 &  \lambda^{6} & \lambda^4\\
 \lambda^{6} & 1 & \lambda^2\\
 \lambda^4  &\lambda^2 & 1
\end{array}
\right).
\end{split}
\end{equation}
The trilinear soft SUSY breaking terms have a similar structure to the Yukawa matrices due to the horizontal symmetry,
but they are not exactly proportional to the Yukawa matrices in general due to their $\mathcal{O}(1)$ undetermined coefficients.

So far, we have implicitly assumed the kinetic terms of quarks and leptons are canonical, but in general,
their K{\"a}hler potential can be
\begin{align}
    K_{\rm matter} =  X^\dagger_i Z_{ij} X_j\ ,
\end{align}
where $X_i = Q_i, \bar u_i, \bar d_i, L_i, \bar \nu_i$ or $\bar e_i$,
and $Z_{ij}$ is a Hermitian matrix.
The off-diagonal elements of $Z_{ij}$ are suppressed by some powers of $\lambda$, following their horizontal charges.
To obtain the canonical kinetic terms, we diagonalize $Z_{ij}$
by rotating the field as $X_i=V_{ij}^X X'_j$ with a unitary matrix $V^X$
and further redefine $X'_j$ to remove the remaining $\mathcal{O}(1)$ numbers in the diagonal parts. 
This process can change the orders of magnitude of some entries of the Yukawa and soft mass-squared matrices
in the basis of the canonically normalized kinetic terms. (Because we did not take all $U(1)_H$ charges to have the same sign, the argument of Appendix B of ref.~\cite{Leurer:1993gy} does not apply.)
We take account of this effect in our numerical analyses.

The $U(1)_H$ horizontal symmetry is anomalous with the charge assignments presented above.
The anomalies arise from
    $U(1)_Y^2-U(1)_H$,~$U(1)_H^2-U(1)_Y$,~$U(1)_H^3$,~
    $U(1)_H-({\rm Gravity})^2$,~$U(1)_H-SU(3)_C^2$ and $U(1)_H-SU(2)_L^2$.
While the anomaly cancellation may be realized by the 4d Green-Schwarz mechanism \cite{Green:1984sg, Witten:1984dg},
instead the anomalies could be canceled by adding new fields charged under the horizontal symmetry.
Another possibility is to find an anomaly-free discrete group $\mathbb{Z}_N\subset U(1)_H$. 
If $N$ is large enough, like $N\geq 20$, the powers of $\lambda$ for the Yukawa and soft mass-squared matrices
are not affected.

\subsection{$R$-symmetry and the $\mu$ problem}
\label{subsec:RsymMu}

Throughout this paper, we follow~\cite{Aloni:2021wzk} in assuming a $\mathbb{Z}_4$ $R$-symmetry~\cite{Lee:2010gv}, under which the SM matter fields have charge $1$ and the Higgs fields, the flavons, and the SUSY-breaking spurions have charge $0$. Lagrange multiplier fields like $X$ and $Y$ in \eqref{eq:CPVsuper1} carry charge  $2$. This symmetry has the effect of forbidding a superpotential $\mu$-term, but allowing the Giudice-Masiero mechanism to generate effective $\mu$ and $b_\mu$ terms from K\"ahler potential terms~\cite{Giudice:1988yz}. We assume that the SUSY-breaking spurions do not violate flavor or CP symmetries. As a result, the CP phase of the $\mu$ term arises only from higher order, flavon-suppressed terms. For example, in the model discussed above, one expects to generate a $\mu$ term  of order
\begin{equation}
\mu \sim m_{3/2} \left(1 + {\cal O}({\bar S}_2^3 S_3^2)\right) \sim m_{3/2}\left(1 + i {\cal O}(\lambda^{12})\right).
\end{equation}
Here  ${\bar S}_2^3 S_3^2$ is the leading flavor-invariant but CP-violating  term that can be constructed from the flavon fields. This model predicts that EDM contributions sensitive to $\arg \mu$ are extremely suppressed, by a factor $\lambda^{12} \approx 4 \times 10^{-9}$ relative to naive expectations.

\subsection{CP and flavor bounds}

Let us now discuss CP and flavor constraints on the model presented in the previous subsection.
The most relevant observables are summarized in Tab.~\ref{tab:flavor_constraint}.
We estimate SUSY contributions to these observables by using the mass insertion approximation.
As a demonstration, we take $\widetilde m_q = M_3=20$\,TeV,
$\widetilde m_\ell=M_{1,2}= \mu=10$\,TeV and $\tan\beta=50$
where $M_{1,2,3}$ are three gaugino masses.
The typical mass scales of the trilinear soft SUSY breaking terms are taken as $\widetilde m_q,~\widetilde m_\ell$
for squarks and sleptons, respectively.
We have checked that the estimation is consistent with that of the public code, {\sf susy\_flavor\_v2.5}~\cite{Rosiek:2010ug,Crivellin:2012jv,Rosiek:2014sia}.

The CP-violating parameter $\epsilon_K$ measured from the neutral Kaon oscillation
is generally sensitive to supersymmetric particles
whose masses are much higher than the TeV scale
(see, e.g., ref.~\cite{Altmannshofer:2013lfa}).
The current experimental value is
$
   |\epsilon_K|=  2.228(11)\times 10^{-3}
$\
\cite{Tanabashi:2018oca}.
The theoretical uncertainty of the SM prediction is, however, more than one order of magnitude larger than
the experimental value due to the large uncertainty of $|V_{cb}|$~\cite{Kim:2019vic}.
In the model with three flavons, the SUSY contribution to $\epsilon_K$ is estimated as~\cite{Ciuchini:1998ix}
\begin{align}
\label{epsilonK}
    |\epsilon_{K}^{\rm SUSY}|\simeq 
   10^2 \left(\frac{20\,{\rm TeV}}{\widetilde m_q}\right)^2{\rm Im}[(\delta^d_{12})_{LL}(\delta^d_{21})_{RR}]
    \ .
\end{align}
Here, $(\delta^X_{ij})_{HH}$ is defined through the soft mass-squared matrix for the scalar partner of a SM fermion $X$
in the Yukawa diagonal basis,
$(\hat M^2_{\widetilde X})_{ij} \equiv \widetilde m^2(\delta_{ij}+(\delta^X_{ij})_{HH})$,
where $H$ denotes the helicity of the fermion $X$.
To obtain the soft mass-squared matrices or the trilinear soft SUSY breaking parameters in the Yukawa diagonal basis, we use a method similar to the one discussed in ref.~\cite{Aloni:2021wzk}.
We introduce a random parameter for each entry of the flavor matrices and find parameter sets
which satisfy criteria to fit with the observed fermion masses, mixing angles and CP phases.
Finding $1000$ data sets to satisfy the criteria, we compute, e.g., averaged soft mass-squared matrices.
See appendix~\ref{app:criteria} for more details.
By using this method, we obtain  $(\delta^d_{12})_{LL}(\delta^d_{21})_{RR}\approx \lambda^{6.7}$,
for the averaged values, which lead to $|\epsilon_{K}^{\rm SUSY}|\approx 10^{-3}$ with an $\mathcal{O}(1)$ phase.
This contribution is comparable to the observed value.

For the $D-\bar D$ mixing, the current experimental value of the mass difference
is given by
$
|{\mit \Delta}M_D| = 0.63^{+0.27}_{-0.29}\times 10^{-14}\,{\rm GeV}
$~\cite{Tanabashi:2018oca}.
The uncertainty of the SM prediction is also expected to be large due to long distance effects
(see, e.g., ref.~\cite{Amhis:2012bh} and references therein).
The SUSY contribution to $\Delta M_{D}$ is given by~\cite{Gabbiani:1996hi,Altmannshofer:2009ne}
\begin{align}
\begin{split}
    |\Delta M_{D}^{\rm SUSY}| \simeq \,\, 10^{-12} \, {\rm GeV} &\times \left(\frac{20\,{\rm TeV}}{\widetilde m_q}\right)^2 \\
    &\times {\rm Re}\left[
    (\delta^u_{12})_{LL}(\delta^u_{21})_{RR}\right] ,
\end{split}    
\end{align}
where 
we have found $(\delta^u_{12})_{LL}(\delta^u_{21})_{RR}\approx \lambda^{3.5}$ for the averaged values
by using the method described above
and obtain $|\Delta M_{D}^{\rm SUSY}| \approx 10^{-15}\,{\rm GeV}$
comparable to the observed value.

The neutron EDM is also used to put a constraint on flavor mixings and CP phases in the (s)quark sector.
The current limit is given by
$
   |d_n|<1.8\times 10^{-26}~e\,{\rm cm}
$
\cite{Abel:2020gbr}.
The SUSY contribution to the neutron EDM is~\cite{Pokorski:1999hz}
\begin{align}
\begin{split}
\label{neutronEDM}
|d_n^{\rm SUSY}|  
\simeq \,\,  10^{-23} \, e\,{\rm cm} &\times \left(\frac{20\,{\rm TeV}}{\widetilde m_q}\right)^3\left(\frac{M_3}{20\,{\rm TeV}}\right)\\
&\times\,
{\rm Im}\left[\frac{(A^u_{11})_{LR}}{\widetilde m_q}\right] ,
\end{split}    
\end{align}
where $(A^u_{ij})_{LR}$ denotes the $(i,j)$ entry of the trilinear soft SUSY breaking term of the up-type squarks.
Using the averaged value of $(A^u_{11})_{LR}\approx \lambda^{6.2}$
with an $\mathcal{O}(1)$ phase,
we find  $|d_n^{\rm SUSY}| \approx 10^{-28} \, e\,{\rm cm}$.

Among LFV processes, ${\rm Br}(\mu\to e+\gamma)$ gives the most stringent bound on the SUSY parameter space.
The current upper bound on this process is ${\rm Br}(\mu\to e+\gamma) < 4.2\times  10^{-13}$~\cite{TheMEG:2016wtm}.
The SUSY contribution to ${\rm Br}(\mu\to e+\gamma)$ is estimated in the mass insertion approximation as~\cite{Paradisi:2005fk}
\begin{align}
\begin{split}
    {\rm Br}(\mu\to e+\gamma) \simeq \,5&\times 10^{-9}\left(\frac{\tan\beta}{50}\right)^2
    \left(\frac{10\,{\rm TeV}}{\widetilde m_\ell}\right)^4 \\
    &\times \left(\frac{\mu\, M_1}{\widetilde m_\ell^2}\right)^2
    \left|(\delta^{\ell}_{23})_{LL}(\delta^{\ell}_{31})_{RR}\right|^2 .
\end{split}    
\end{align}
Using the averaged value $|(\delta^{\ell}_{23})_{LL}(\delta^{\ell}_{31})_{RR} |\approx \lambda^{3.3}$, we obtain $ {\rm Br}(\mu\to e+\gamma)\approx\,10^{-13}$, comparable to the current experimental bound.

The electron EDM is also an important probe for flavor mixings and CP phases in the (s)lepton sector.
The current upper limit is given by $|d_e| < 1.1\times 10^{-29}~e$\,cm~\cite{Andreev:2018ayy}.
The SUSY contribution to the electron EDM is
\begin{align}
\begin{split}
\label{electronEDM}
    |d_e^{\rm SUSY}|\simeq 5 &\times 10^{-25}\,e\,{\rm cm} \times \left(\frac{\tan\beta}{50}\right)\left(\frac{10\,{\rm TeV}}{\widetilde m_\ell}\right)^2\\
    &\times \left(\frac{\mu\,M_1}{\widetilde m_\ell^2}\right) {\rm Im}[(\delta^\ell_{13})_{LL}(\delta^\ell_{31})_{RR}]\ .
\end{split}    
\end{align}
(There are additional contributions that are independent of flavor violation but proportional to $\arg \mu$, but these are more suppressed because the phase of $\mu$ is very small in this model, as discussed above in section~\ref{subsec:RsymMu}.)
Using the averaged value $|(\delta^\ell_{13})_{LL}(\delta^\ell_{31})_{RR}|\approx \lambda^{6.8}$ with an $\mathcal{O}(1)$ phase,
we obtain $|d_e^{\rm SUSY}|\approx 5\times 10^{-30}\,e\,{\rm cm}$, which is comparable to the current limit.

\begin{table*}[ht]
\centering

\begin{tabular}{|c|c|c|c|c|}
\hline 
$\quad$ Observable $\quad$ & $\qquad$ Experimental bound $\qquad$ & $\quad$ Model with $S_3$ $\quad$ & $\quad$ Model with $S_4$ $\quad$ \\ 
\hline 
$|\epsilon_K|$  &  $2.228(11)\times 10^{-3}$ \cite{Tanabashi:2018oca} & $\sim 10^{-3} $ & $\sim 10^{-7}$ \\ 
\hline 
$|{\mit \Delta}M_D|$  & $0.63^{+0.27}_{-0.29}\times 10^{-14}\,{\rm GeV}$ \cite{Tanabashi:2018oca} & $\sim 5\times 10^{-15}\,{\rm GeV}$
& $\sim 5\times 10^{-15}\,{\rm GeV}$ \\ 
\hline 
nEDM  & $\leqslant 10^{-26}~e$\,cm \cite{Abel:2020gbr} & $\sim 10^{-28}~e$\,cm & $\sim 10^{-28}~e$\,cm \\ 
\hline 
${\rm Br}(\mu\to e+\gamma)$  & $\leqslant 4.2\times  10^{-13}$~\cite{TheMEG:2016wtm} &  $\sim 10^{-16}$
& $\sim 10^{-16}$ \\
\hline
eEDM  & $\leqslant 1.1\times 10^{-29}~e$\,cm~\cite{Andreev:2018ayy} & $\sim 5\times  10^{-29}~e$\,cm & $\sim 10^{-30}~e$\,cm  \\
\hline
\end{tabular}
\caption{CP and flavor observables and their current experimental bounds.
The estimation of the SUSY contribution to each observable in the models
with $U(1)_{H_1} \times U(1)_{H_2}$ is also shown.
We take $\widetilde m_q = M_3 =5$\,TeV, $\widetilde m_\ell = M_{1,2} = \mu = 500\,{\rm GeV}$ and $\tan\beta=50$.
The typical mass scales of the trilinear soft SUSY breaking terms are taken as $\widetilde m_q,~\widetilde m_\ell$
for squarks and sleptons, respectively.}
\label{tab:flavor_constraint}
\end{table*}

%#######################
\section{Models with two $U(1)$ symmetries}\label{viableModel}

The model presented in the previous section can accommodate sleptons whose typical mass scale is
required to be around $10$ TeV by CP and flavor constraints.
On the other hand, the muon $g-2$ anomaly can be addressed by light sleptons and electroweakinos
whose masses are less than about 1 TeV.
Here, we explore SUSY models with two $U(1)$ symmetries, $U(1)_{H_1} \times U(1)_{H_2}$,
to pursue the possibility of explaining the muon $g-2$ anomaly without violating CP and flavor constraints.
The symmetries are spontaneously broken by flavon chiral superfields, $S_1$ and $S_2$,
which are singlets under the SM gauge symmetry.
Their charges under $U(1)_{H_1} \times U(1)_{H_2}$ are
\begin{equation}
\begin{split}
\label{eq:mass_ratio}
&\begin{array}{c}
S_1 \\
(-1, 0)
\end{array}
\quad 
\begin{array}{c}
S_2 \\
(0, -1)
\end{array}
\end{split}
\end{equation}
and their VEVs are assumed to be $\langle S_1 \rangle \sim \lambda$ and $\langle S_2 \rangle \sim \lambda^2$.

Quarks and leptons carry charges $(H_1,H_2)$ under the $U(1)$ symmetries.
Defining $H = H_1 + 2 H_2$,
the SM fermion mass ratios in Eq.~\eqref{eq:quark_mass} and Eq.~\eqref{eq:lepton_mass}
impose the following constraints on charge assignments for (s)quarks and (s)leptons under the horizontal symmetries, 
\begin{equation}
\begin{split}
\label{eq:mass_ratio_2}
&H(Q_3)+H(\bar u_3)=0, \quad H(Q_2)+H(\bar u_2)=3 , \\[1ex]
&H(Q_1)+H(\bar u_1)=7,\\[1ex]
&H(Q_3)+H(\bar d_3)=0, \quad H(Q_2)+H(\bar d_2)=2,\\[1ex]
&H(Q_1)+H(\bar d_1)=4,\\[1ex]
&H(L_3)+H(\bar e_3)=0, \quad H(L_2)+H(\bar e_2)=2,\\[1ex]
&H(L_1)+H(\bar e_1)=5.
\end{split}
\end{equation} 
We have assumed $H(H_u) = H(H_d) = 0$, and
the constraints work well for a large $\tan\beta\sim 50$.
Using Eq.~\eqref{CKMPMNS}, the constraints from the CKM and PMNS mixing matrices in Eq.~\eqref{ckm} and Eq.~\eqref{pmns}
are expressed as
\begin{equation}
\begin{split}
\label{eq:CKM_PMNS}
&H(Q_1)-H(Q_2)=1, \quad H(Q_1)-H(Q_3)=3,\\[1ex]
&H(Q_2)-H(Q_3)=2,\\[1ex]
&H(L_1)-H(L_2)=H(L_1)-H(L_3)=1, \\[1ex]
&H(L_2)-H(L_3)=0.
\end{split}
\end{equation}
Many models can satisfy the conditions of Eq.\,\eqref{eq:mass_ratio_2} and Eq.\,\eqref{eq:CKM_PMNS}.
Here, we present a working example of possible charge assignments for (s)quarks and (s)leptons,
\begin{equation}
\begin{split}
&\begin{array}{c}
Q_1 \\
(3, 0)
\end{array}
\,\,
\begin{array}{c}
Q_2 \\
(0, 1)
\end{array}
\,\,
\begin{array}{c}
Q_3 \\
(0, 0)
\end{array}
\quad 
\begin{array}{c}
\bar{u}_1 \\
(-2, 3)
\end{array}
\,\,
\begin{array}{c}
\bar{u}_2 \\
(1, 0)
\end{array}
\,\,
\begin{array}{c}
\bar{u}_3 \\
(0, 0)
\end{array} \\[1ex]
&\begin{array}{c}
\bar{d}_1 \\
(-3, 2)
\end{array}
\,\,
\begin{array}{c}
\bar{d}_2 \\
(2, -1)
\end{array}
\,\,
\begin{array}{c}
\bar{d}_3 \\
(0, 0)
\end{array} \\[1ex]
&\begin{array}{c}
L_1 \\
(5, 0)
\end{array}
\,\,
\begin{array}{c}
L_2 \\
(0, 2)
\end{array}
\,\,
\begin{array}{c}
L_3 \\
(0, 2)
\end{array}
\quad 
\begin{array}{c}
\bar{e}_1 \\
(-4, 2)
\end{array}
\,\,
\begin{array}{c}
\bar{e}_2 \\
(2, -2)
\end{array}
\,\,
\begin{array}{c}
\bar{e}_3 \\
(0, -2)
\end{array}
\label{twohorizontalcharges}
\end{split}
\end{equation}

As in the case of the model with a single $U(1)_H$,
the Yukawa matrices are given by couplings with the flavons $S_1, S_2$
and their orders of magnitude can be parameterized by $\lambda$,
\begin{equation}
\begin{split}
\label{eq:q_yukawa_extended_2}
&Y_u\sim
\left(
\begin{array}{ccc}
\lambda^7 &  \lambda^4 & \lambda^3 \\
0  & \lambda^3 & \lambda^2\\
0 & \lambda & 1
\end{array}
\right), \quad \,\, Y_d\sim
\left(
\begin{array}{ccc}
\lambda^4 & 0 & \lambda^3\\
0 &\lambda^2 & \lambda^2\\
0 & 0 & 1
\end{array}
\right),\\[1ex]
&Y_e\sim
\left(
\begin{array}{ccc}
\lambda^5 & 0 & 0\\
0 & \lambda^2 & 1\\
0 & \lambda^2 & 1
\end{array}
\right), \qquad Y_\nu\sim
\left(
\begin{array}{ccc}
\lambda^{10} & \lambda^9 & \lambda^9\\
\lambda^9 & \lambda^8 & \lambda^8 \\
\lambda^9 & \lambda^8 & \lambda^8
\end{array}
\right).
\end{split}
\end{equation}
One important difference from the case of a single $U(1)_H$ is that
some entries are zero because of the holomorphic nature of superpotential terms~\cite{Seiberg:1993vc,Leurer:1993gy}.
For example, if $H_1(Q_i) + H_1(\bar{u}_j) <0$ or $H_2(Q_i) + H_2(\bar{u}_j) <0$, then $(Y_u)_{ij}$ vanishes.
In the previous model with a single $U(1)_H$, the flavon field $\bar{S}_2$ with a positive charge was introduced,
while the current model does not have such a flavon field.

Spontaneous CPV is not realized with only the two flavons $S_1$, $S_2$
because phases of their VEVs are rotated away by the $U(1)_{H_1} \times U(1)_{H_2}$.
Then, we introduce an additional singlet chiral superfield $S_N$ where $N\geq 3$ is a positive integer.
Its horizontal charges are $(-N,0)$
and the VEV, $\langle S_N \rangle \sim \lambda^N$, is complex.
Such a VEV is easily obtained by considering superpotential terms,
\begin{equation}
\begin{split}
W_S = Z ( a S_N^2 + b S_N S_1^N + c S_1^{2N} ) \, , 
\end{split}
\end{equation}
where a singlet chiral superfield $Z$ has horizontal charges $(2N,0)$.
The $F$-term conditions $F_{S_1} = F_{S_N} = 0$ are solved by $\langle Z \rangle = 0$,
and $F_{Z} = 0$ leads to
\begin{equation}
\begin{split}
a \langle S_N \rangle^2 + b \langle S_N \rangle \langle S_1 \rangle^N + c \langle S_1 \rangle^{2N} = 0 \, .
\end{split}
\end{equation}
This equation is solved as
\begin{equation}
\begin{split}
\frac{\langle S_N \rangle}{\langle S_1 \rangle^N} = \frac{-b \pm \sqrt{b^2 -4ac}}{2a} \, .
\end{split}
\end{equation}
We take $b^2 -4ac < 0$ to get a complex $\langle S_N \rangle$.
Through superpotential couplings with $S_N$, CP phases are introduced to the Yukawa matrices.
For $N=3$, for example, $(Y_d)_{13}$ receives a new contribution with $|\langle S_3 \rangle| \sim \lambda^3$,
which leads to an $\mathcal{O}(1)$ phase of the CKM matrix.
For $N=4$, only the $(Y_u)_{12}$ obtains a phase, which also leads to an $\mathcal{O}(1)$ CKM phase.
For $N\geq 5$, the quark Yukawa matrices do not receive $\mathcal{O}(1)$ CKM phases,%
\footnote{Contributions from the non-canonical kinetic terms to the CKM phase are negligible.}
so we focus on the cases of $N=3,4$ in the following discussion.
For the lepton sector, an $\mathcal{O}(1)$ phase is also introduced to $Y_\nu$.  Notice that the flavor-invariant, CP-violating combination that can contribute to $\arg \mu$ (or other flavor-invariant terms originating in the K\"ahler potential) is $S_1^N S_N^\dagger$, of order $\lambda^{2N}$. Thus, the $\arg \mu$ contribution to EDMs can be significantly larger than that discussed in section~\ref{subsec:RsymMu}.

With the charge assignments of Eq.~\eqref{twohorizontalcharges},
the structure of the soft scalar mass-squared matrices for squarks and sleptons is constrained as
\begin{equation}
\begin{split}
\label{eq:squark_msq_2}
&M^2_{\widetilde Q}\sim\widetilde m^2
\left(
\begin{array}{ccc}
1 & \lambda^5 & \lambda^{3} \\
\lambda^5  & 1 & \lambda^2\\
\lambda^3 & \lambda^2 & 1
\end{array}
\right) ,
\,\, M^2_{\widetilde {\bar u}}\sim\widetilde m^2
\left(
\begin{array}{ccc}
1  & \lambda^9 &  \lambda^8\\
 \lambda^9 & 1 & \lambda\\
 \lambda^8 & \lambda & 1
\end{array}
\right),\\[1ex]
&M^2_{\widetilde{\bar d}} \sim \widetilde m^2
\left(
\begin{array}{ccc}
1 &  \lambda^{11} &\lambda^{7}\\
 \lambda^{11} & 1 &\lambda^4\\
\lambda^{7} & \lambda^4 & 1
\end{array}
\right),\\[1ex]
&M^2_{\widetilde L}\sim\widetilde m^2
\left(
\begin{array}{ccc}
1 & \lambda^9 &  \lambda^9 \\
 \lambda^9 & 1 & 1\\
 \lambda^9 & 1 & 1
\end{array}
\right), \,\, M^2_{\widetilde{\bar e}}\sim\widetilde m^2
\left(
\begin{array}{ccc}
1 &  \lambda^{14} & \lambda^{12}\\
 \lambda^{14} & 1 & \lambda^2\\
 \lambda^{12}  &\lambda^2 & 1
\end{array}
\right).
\end{split}
\end{equation}
The trilinear soft SUSY breaking terms have the same structure as the Yukawa matrices,
up to $\mathcal{O}(1)$ undetermined coefficients.

As in the case of the model with a single $U(1)_H$,
we take account of the effect of the non-canonically normalized kinetic terms for
(s)quarks and (s)leptons in the following analysis.
The $U(1)_{H_1} \times U(1)_{H_2}$ horizontal symmetries are also anomalous. 
The same comment can be applied to the present model.
We have not provided a concrete mechanism to generate VEVs
$\langle S_1 \rangle \sim \lambda$ and $\langle S_2 \rangle \sim \lambda^2$.
In the Green-Schwarz mechanism, the Fayet-Iliopoulos (FI) term is generated
for an anomalous $U(1)$ via the gravitational anomaly~\cite{Dine:1987xk,Atick:1987gy},
which leads to a nonzero flavon VEV~\cite{Babu:2009fd}.
This mechanism may be able to be applied to the present model, which is left for a future study.

.

\begin{figure*}
\begin{minipage}[t]{\hsize}
\includegraphics[width=6cm]{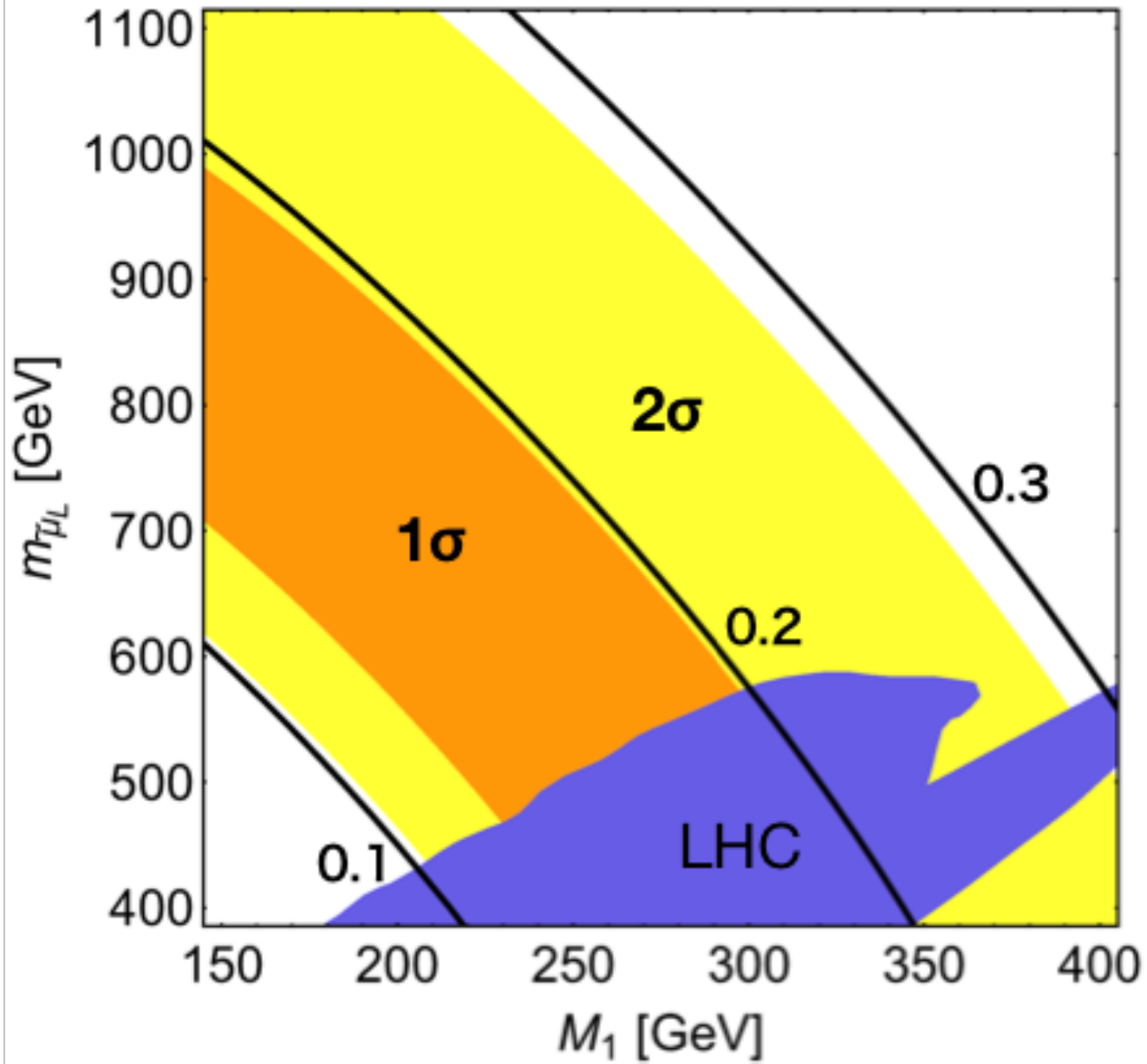}
\hspace{2cm}
\includegraphics[width=6cm]{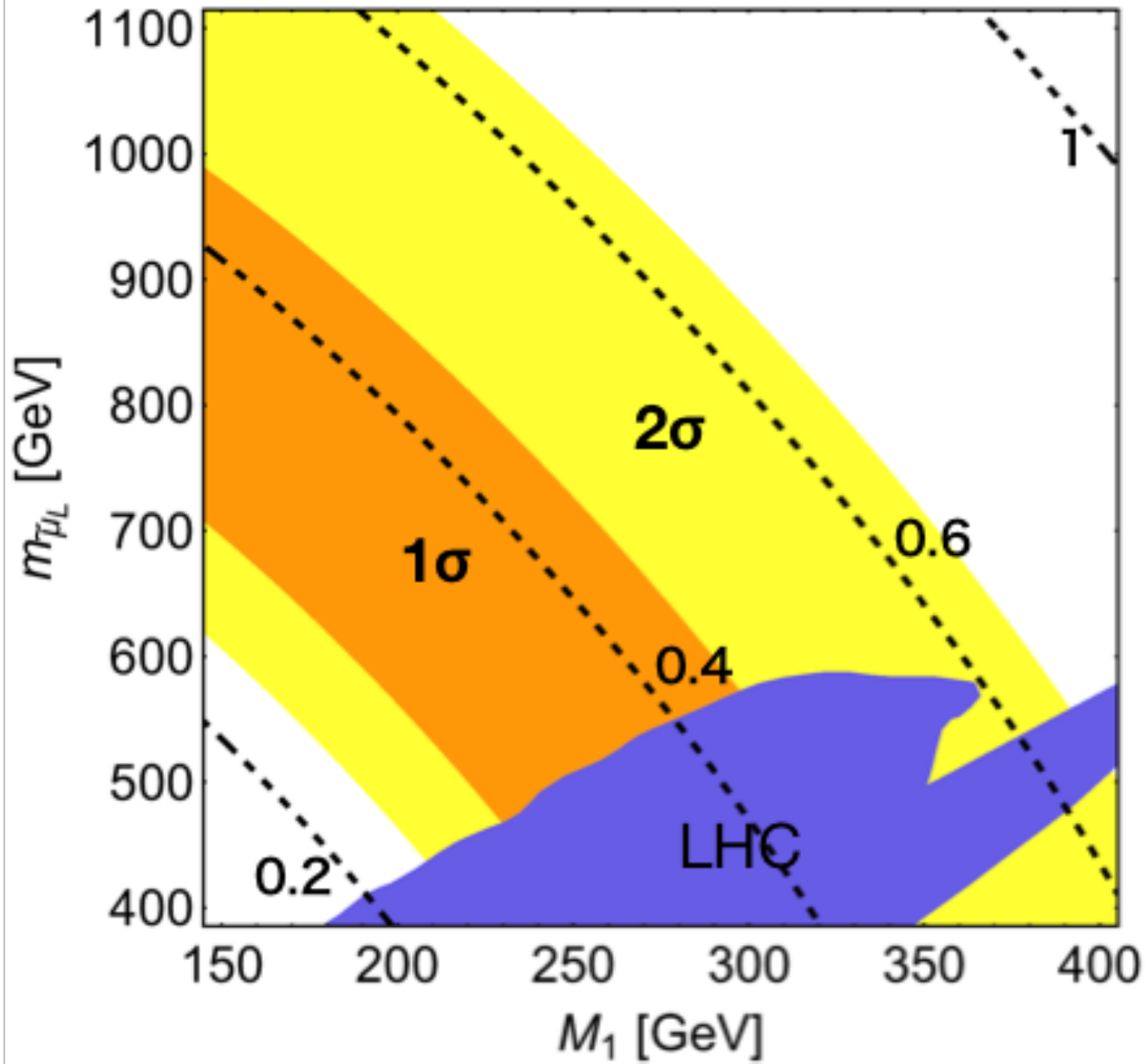}
  \end{minipage}
  \caption{ The muon $g-2$ anomaly can be addressed by the SUSY contribution at the $1\sigma$ ($2\sigma$) level
in the orange (yellow) shaded region. The left and right panels correspond to the $U(1)_{H_1}\times U(1)_{H_2}$ models
with $S_3$ and $S_4$, respectively.
In both panels, we assume only the left-handed sleptons and electroweakinos (bino, wino and higgsinos) are light
and the right-handed sleptons, all squarks, the gluino, and the heavy Higgs bosons are decoupled.
We take $\tan\beta=50$ and $\mu=M_2=2M_1$.
To estimate the SUSY contribution to the muon $g-2$,
the public code {\sf susy\_flavor\_v2.5}~\cite{Rosiek:2010ug,Crivellin:2012jv,Rosiek:2014sia} is used.
The result is consistent with the one-loop calculation presented in ref.~\cite{Moroi:1995yh}. 
The purple shaded region is excluded by searches for slepton pair production
at the Large Hadron Collider (LHC)~\cite{Sirunyan:2017lae,Aaboud:2018jiw,Aad:2019vnb,Sirunyan:2020eab},
as discussed in refs.~\cite{Endo:2020mqz,Endo:2021zal}.
The black solid lines in the left panel correspond to the current electron EDM bound, $|d_e|=1.1\times 10^{-29}\,e\,{\rm cm}$,
by using $\mu=|\mu|(1+i\,\kappa \lambda^6)$ where $\lambda=0.2$ and $\kappa=0.1,~0.2,~0.3$
from the left to the right, respectively.
The black dashed lines in the right panel correspond to the future reach of electron EDM measurements,
$|d_e|=10^{-30}\,e\,{\rm cm}$, by using $\mu=|\mu|(1+i\,\kappa\,\lambda^8)$ where $\lambda=0.2$ and $\kappa=0.2,~0.4,~0.6,~1$
from the left to the right, respectively.
}
\label{fig:muon_g2} 
\end{figure*}

%#######################
\section{Flavor constraints \& $(g-2)_\mu$}\label{flavorconst}

We now discuss CP and flavor constraints
on the model with two $U(1)$ horizontal symmetries presented in the previous section
and see if the model can successfully address the muon $g-2$ anomaly. 
Here, for reference values of the model parameters, we take $\widetilde m_q = M_3 =5$\,TeV,
$\widetilde m_\ell = M_{1,2} = \mu = 500\,{\rm GeV}$ and $\tan\beta=50$.
The typical mass scales of the trilinear soft SUSY breaking terms are also taken as $\widetilde m_q,~\widetilde m_\ell$
for squarks and sleptons, respectively.
The observed Higgs mass can be explained with the stop trilinear soft SUSY breaking parameter for a squark mass around $5\,{\rm TeV}$
\cite{Draper:2011aa}.
We have checked the consistency of the estimates with those obtained by the public code
{\sf susy\_flavor\_v2.5}.
The SUSY contributions to the quark and lepton sector observables are 
summarized in Tab.~\ref{tab:flavor_constraint} and generally suppressed
compared to those of the model with a single $U(1)$ due to the further suppression of
the off-diagonal entries of the Yukawa and soft mass-squared matrices.

Let us first discuss the quark sector observables.
Here again, $1000$ good trials to realize the observed fermion masses, mixing angles
and CP phases are found, and the soft mass-squared matrices are averaged over the trials
to compute the observables.
In the present model, the most dominant contribution to $\Delta M_{D}$ is given by~\cite{Gabbiani:1996hi,Altmannshofer:2009ne}
\begin{equation}
\begin{split}
    |\Delta M_{D}^{\rm SUSY}| \simeq \,\, &10^{-11} \, {\rm GeV} \times \left(\frac{5\,{\rm TeV}}{\widetilde m_q}\right)^2 \\
    &\times \left|{\rm Re}\left[
    (\delta^u_{12})_{LL}(\delta^u_{21})_{RR}\right]\right| .
\end{split}    
\end{equation}
Here, the averaged values of $(\delta^u_{12})_{LL}(\delta^u_{21})_{RR}\approx\lambda^{4.8}$
lead to $|\Delta M_{D}^{\rm SUSY}|\approx 5\times 10^{-15} \, {\rm GeV}$, comparable to the experimental value.
The SUSY contribution to $\epsilon_K$ given by Eq.~\eqref{epsilonK} is quite different between the models with $S_3$ and $S_4$.
For the former case, by using the averaged values of  $(\delta^d_{12})_{LL}(\delta^d_{21})_{RR}\approx \lambda^8$
with $\mathcal{O}(1)$ phases,
we obtain $|\epsilon_{K}^{\rm SUSY}|\approx 10^{-3}$,
which is comparable to the observed value.
On the other hand, for the latter case, the SUSY contribution is quite suppressed as $|\epsilon_{K}^{\rm SUSY}|\approx 10^{-7}$ with ${\rm Im}[(\delta^d_{12})_{LL}(\delta^d_{21})_{RR}]\approx \lambda^{14}$. This is because the flavon $S_4$ has a larger $U(1)_{H_1}$ charge compared to that of $S_3$ and CP phases are provided for fewer entries of the mass matrices of, in particular, the down (s)quark sector.
The SUSY contribution to the neutron EDM is presented in Eq.~\eqref{neutronEDM}.
Using the averaged value of ${\rm Im}[(A^u_{11})_{LR}/\widetilde m_q]\approx \lambda^{8}$, we obtain $|d_n^{\rm SUSY}|\approx 10^{-28}\,e\,{\rm cm}$, smaller than the current limit.

For the lepton sector observables, the dominant contribution to ${\rm Br}(\mu\to e+\gamma)$
is now estimated in the mass insertion approximation as~\cite{Paradisi:2005fk}
\begin{align}
\begin{split}
    {\rm Br}(\mu\to e+\gamma) \simeq \,\, &5\times 10^{-4}\left(\frac{\tan\beta}{50}\right)^2\\
    &\times \left(\frac{500\,{\rm GeV}}{\widetilde m_\ell}\right)^4 \left(\frac{\mu\, M_1}{\widetilde m_\ell^2}\right)^2\\
    &\times 
     \left(\left|0.5(\delta^{\ell}_{21})_{LL}\right|^2+\left|(\delta^{\ell}_{23})_{RR}(\delta^{\ell}_{31})_{LL}\right|^2\right).
\end{split}    
\end{align}
Here, the averaged values are given by $(\delta^{\ell}_{21})_{LL}\approx \lambda^{8.7}$,
$(\delta^{\ell}_{23})_{RR}(\delta^{\ell}_{31})_{LL}\approx \lambda^{9.7}$,
which lead to $ {\rm Br}(\mu\to e+\gamma)\approx 10^{-16}$.
The contribution is much smaller than the experimental upper bound.

The SUSY contribution to the electron EDM from the off-diagonal entries of the SUSY breaking parameters is negligible,
and the dominant contribution is obtained from the flavor diagonal entries with the phase of $\mu$.
Then, the contribution to the electron EDM is related to that of the muon $g-2$,
\begin{equation}
\begin{split}
    |d_e^{\rm SUSY}|&\sim e\frac{m_e}{m_\mu}\frac{a_\mu}{2m_\mu} |{\rm arg}(\mu)|\\
    &\simeq  10^{-24}\left(\frac{a_\mu}{2\times 10^{-9}}\right)|{\rm arg}(\mu)|\,e\,{\rm cm}\ .
\end{split}
\end{equation}
This relation shows that a tiny phase of $\mu$  is required to explain the muon $g-2$ anomaly
and to be consistent with the current upper bound on the electron EDM,
i.e., $|{\rm arg}(\mu)|\lesssim 10^{-5}$ for $a_\mu\approx 2\times 10^{-9}$.
The electron EDM is then a powerful probe for SUSY models addressing the muon $g-2$ anomaly.
In the model with $S_3$, we estimate ${\rm arg}(\mu)\sim \lambda^6$
and obtain $|d_e^{\rm SUSY}|\simeq 5\times 10^{-29}\,e\,{\rm cm}$ for $a_\mu\simeq 2\times 10^{-9}$,
which requires an $\mathcal{O}(10)\%$ fine-tuning to be consistent with the current electron EDM bound.
The model with $S_4$ gives $|{\rm arg}(\mu)|\sim \lambda^8$
leading to $|d_e^{\rm SUSY}|\simeq 10^{-30}\,e\,{\rm cm}$ for $a_\mu\simeq 10^{-9}$.
Future electron EDM experiments
(see, e.g., refs.~\cite{Kozyryev:2017cwq,Vutha:2017pej,NL-eEDM:2018lno,Ho:2020ucd,Hutzler:2020lmj,Fitch:2021pfs})
will search for the favored parameter space to explain the muon $g-2$ anomaly.

Figure~\ref{fig:muon_g2} shows the parameter space where the muon $g-2$ anomaly can be addressed by the SUSY contribution
in the $U(1)_{H_1}\times U(1)_{H_2}$ models with $S_3$ and $S_4$.
Here, we assume only the left-handed sleptons and electroweakinos (bino, wino and higgsinos) are light
and the right-handed sleptons, all squarks, gluino and heavy Higgs bosons are decoupled.
To estimate the SUSY contribution to the muon $g-2$,
the public code {\sf susy\_flavor\_v2.5}~\cite{Rosiek:2010ug,Crivellin:2012jv,Rosiek:2014sia} is used.
The result is consistent with the one-loop calculation presented in ref.~\cite{Moroi:1995yh}. 
We also take account of searches for slepton pair production
at the LHC~\cite{Sirunyan:2017lae,Aaboud:2018jiw,Aad:2019vnb,Sirunyan:2020eab},
as discussed in refs.~\cite{Endo:2020mqz,Endo:2021zal}.
The black solid lines in the left panel correspond to the current electron EDM constraint $|d_e|=1.1\times 10^{-29}\,e\,{\rm cm}$
with different amounts of tuning in the nonzero CP phase contribution to the $\mu$ parameter.
The black dashed lines in the right panel denote the future reach of electron EDM measurements $|d_e|=10^{-30}\,e\,{\rm cm}$.
As we discussed above, the favored parameter space to explain the muon $g-2$ anomaly without tuning is all covered
by the future measurements.

%#######################
\section{Conclusions and discussions}\label{Conclusion}

SUSY provides an attractive possibility to explain the reported muon $g-2$ anomaly,
but SUSY contributions to LFV and CPV processes must be sufficiently suppressed.
In this paper, we have considered $U(1)$ horizontal symmetries
to address hierarchical masses of quarks and leptons.
Such SUSY alignment models with spontaneous CP violation can also control the structure of sfermion masses
and suppress CP and flavor violating processes.
The correct CKM phase in the quark sector is realized at the same time.
We started with a model with a single $U(1)$ horizontal symmetry
and investigated CP and flavor constraints. 
The model can viably achieve supersymmetric particles at around $10 \, \rm TeV$ for a large $\tan \beta$. 
Then, we considered a model with two $U(1)$ horizontal symmetries.
The model can further relax CP and flavor constraints and 
realize sleptons and electroweakinos at a scale $\lesssim \mathcal{O}(1) \, {\rm TeV}$
to provide a viable solution to the muon $g-2$ anomaly.
We found that the favored parameter space to address the muon $g-2$ anomaly will be extensively investigated by
future electron EDM experiments.

The lepton sector CPV is hinted by the neutrino oscillation,
although $\delta_{\rm CP}=\pi$ is still consistent.
Our models generically predict a sizable CP phase in the lepton sector,
and hence its discovery would support the models. 
Unlike gauge mediation or gaugino mediation of SUSY breaking, the selectron and smuon masses are generally not degenerate
in SUSY alignment models.
If the LHC observes the selectron and smuon with different masses, it may indicate the existence of horizontal symmetries.

%---------------SECTION------------------%
%
\section*{Acknowledgments}

We would like to thank Daniel Aloni and Pouya Asadi for discussions. MR is supported in part by the DOE Grant DE-SC0013607 and the Alfred P.~Sloan Foundation Grant No.~G-2019-12504.

\appendix

\section{Good trials}
\label{app:criteria}

To estimate SUSY contributions to CP and flavor observables,
we need to find the soft mass-squared matrices of squarks and sleptons in the Yukawa diagonal basis
where $Y_u$, $Y_d$ and $Y_e$ are diagonal.
Following the procedure of ref.~\cite{Aloni:2021wzk},
we first introduce an $\mathcal{O}(1)$ random number for each entry of the Yukawa and soft mass-squared matrices.
Then, going to the Yukawa diagonal basis,
if the generated Yukawa matrices satisfy the following criteria to realize the observed pattern of the quark and lepton masses,
the mixing angles and the CKM phase,
we call it a good trial and compute the effective soft mass-squared matrices.
In a good trial, the ratio of a quark mass $m_q$ and the observed value $m^{\rm obs}_q$ is within a factor of $\lambda=0.2$,
that is,
\begin{align}
    \lambda\leq \frac{m_q}{m_q^{\rm obs}}\leq \frac{1}{\lambda}\ .
\end{align}
The absolute value of each entry of the CKM matrix is within the range,
\begin{align}
\left(
\begin{array}{ccc}
 0.8 & 0.11 & 0.002\\
0.11 & 0.8& 0.02\\
 0.004 & 0.02 & 0.8
\end{array}
\right)  
\leq
    |V_{\rm CKM}|
\leq    
\left(
\begin{array}{ccc}
1& 0.44 & 0.008 \\
0.44 & 1& 0.08\\
0.016 & 0.08 & 1
\end{array}
\right) ,
\end{align}
and the absolute value of the CKM phase defined in the standard notation satisfies
\begin{align}
   0.5 \leq|\sin(\delta_{\rm CKM})|\leq 1\ .
\end{align}
The criteria for the lepton sector is the same as that of ref.~\cite{Aloni:2021wzk}.
We get $1000$ good trials among which 
the absolute values of the real and imaginary parts of each soft mass-squared matrix element are averaged respectively. 
Finally, we divide the matrix by the average of the absolute eigenvalues to make the diagonal entries close to one.

\section{The right-handed neutrinos}
\label{app:seesaw}
One way to UV complete the Majorana neutrino mass terms in Eq.~\eqref{eq:superpotential} is
to introduce three right-handed neutrinos $N_{i}$ $(i =1,2,3)$ and consider the superpotential,
\begin{align}
W_\nu =Y_{L_i} Y_{{N}_j} L_i H_u N_{j}-Y_{N_{k}}Y_{N_l} \widetilde{M}_{N} N_{k} N_{l}\ ,
\end{align}
where $Y_{L_i},Y_{N_{i}}$ are given by couplings to flavons and $\widetilde{M}_{N}$
may be regarded as the order parameter for the spontaneously broken $U(1)_{B-L}$ symmetry.
Here, the same indices $i,j,k,l$ are summed over.
Then, the F-term condition of $N_{i}$ leads to
\begin{align}
Y_{N_{j}} N_{j}&=\frac{Y_{L_i} L_i H_u}{2 \widetilde{M}_{N}}\ .
\end{align}
Integrating out the heavy right-handed neutrinos $N_{i}$ by using this relation,
the effective superpotential is obtained as
\begin{align}
\label{majoranamassterms}
W_{\nu,\rm eff}=\frac{(Y_{L_i} L_i H_u)(Y_{L_j} L_j H_u)}{4 \widetilde{M}_{N}}\ .
\end{align}
Defining $Y_{\nu\,ij}/{M_N} \equiv Y_{L_i} Y_{L_j} / (4 \widetilde{M}_{N})$,
we reproduce the Majorana neutrino mass terms in Eq.~\eqref{eq:superpotential}.
Note that the superpotential \eqref{majoranamassterms} does not depend on $Y_{N_{i}}$
and hence horizontal charges of the right-handed neutrinos $N_{i}$ are not relevant in the discussion.

\bibliography{bib}
\bibliographystyle{utphys}

\end{document}